\documentclass[namedreferences,hyperref,optionalrh]{spr-sola}
\usepackage{graphicx}        
\usepackage{color}           

\chardef\us=`\_

\newcommand{\Lyns}{Ly$\alpha$}
\newcommand{\Ly}{Ly$\alpha\;$}
\newcommand{\flrone}{SOL2010-02-12}
\newcommand{\flrones}{SOL2010-02-12 }
\newcommand{\flrtwo}{SOL2012-07-04}
\newcommand{\flrtwos}{SOL2012-07-04 }

\newcommand{\comment}{}
\newcommand{\commentend}{}
\usepackage{tabularx} 
\usepackage{makecell} 
\usepackage[T1]{fontenc} 
\usepackage{anyfontsize} 
\usepackage{booktabs} 

\mathchardef\mhyphen="2D

\graphicspath{{Figures_Mockup}}

\usepackage{geometry}
\begin{document}

\begin{article}
\begin{opening}

\title{Spectral Irradiance Variability in Lyman-Alpha Emission During Solar Flares}

\author[addressref=aff1,corref,email={lmajury01@qub.ac.uk}]{\inits{L.H.}\fnm{Luke}~\lnm{Majury}\orcid{0009-0002-8491-9593}}

\author[addressref=aff1,corref,email={r.milligan@qub.ac.uk}]{\inits{R.O.}\fnm{Ryan}~\lnm{Milligan}\orcid{0000-0001-5031-1892}}

\author[addressref={aff2,aff1},corref,email={elizabeth.butler@noaa.gov}]{\inits{E.C.}\fnm{Elizabeth}~\lnm{Butler}\orcid{0000-0002-0927-4310}}

\author[addressref=aff1,corref,email={hgreatorex01@qub.ac.uk}]{\inits{H.J.}\fnm{Harry}~\lnm{Greatorex}\orcid{0000-0002-5302-0887}}

\author[addressref={aff3,aff4,aff5},corref,email={maria.kazachenko@colorado.edu}]{\inits{M.D.}\fnm{Maria}~\lnm{Kazachenko}\orcid{0000-0001-8975-7605}}

\address[id=aff1]{Astrophysics Research Centre, School of Mathematics and Physics, Queen's University Belfast, University Road, BT7 1NN, Northern Ireland, UK}

\address[id=aff2]{NOAA/Space Weather Prediction Center, Boulder, CO 80305, USA}

\address[id=aff3]{Dept. of Astrophysical and Planetary Sciences, University of Colorado Boulder, 2000 Colorado Avenue, Boulder, CO 80305, USA}

\address[id=aff4]{National Solar Observatory, 3665 Discovery Drive, Boulder, CO 80303, USA}

\address[id=aff5]{Laboratory for Atmospheric and Space Physics, University of Colorado, Boulder, CO 80303, USA}


\runningauthor{Majury et al.}
\runningtitle{Spectral Irradiance Variability in Lyman-Alpha Emission
During Solar Flares Using SORCE/SOLSTICE}

\begin{abstract}
    The Lyman-alpha (\Lyns; $1216\;\textrm{Å}$) line is the brightest emission line in the quiescent solar spectrum and radiates a significant fraction of the available nonthermal energy during flares. Despite its importance, there is a lack of detailed studies of \Ly spectral variability during flares. Recently, spectrally resolved \Ly flare observations from the SORCE/SOLSTICE instrument have become available.
    This study examines \Ly spectral variability and its relationship with HXR emission from nonthermal electrons, using observations of two M-class flares from SORCE/SOLSTICE and RHESSI. Imaging observations from STEREO/SECCHI EUVI and SDO/AIA provide further context.
    Enhancements across the \Ly line profile were found to closely correlate with bursts of HXR emission, suggesting a primarily nonthermal origin. Red enhancement asymmetries at the peak of each flare were attributed to chromospheric evaporation, while blue wing enhancement and blue asymmetry were linked to a bright filament-eruption seen in SDO/AIA $1600\;\textrm{Å}$ images.
    These findings contribute to the understanding of spectral \Ly variability during flares and highlight the need for future studies using a higher quality and quantity of spectral \Ly flare observations. Such studies will further characterise the physical mechanisms driving \Ly flare variability.
    
\end{abstract}

\end{opening}

\section{Introduction}
    \label{S-Introduction} 
    
    Lyman-alpha (\Lyns) is an \comment{}ultraviolet (UV) line\commentend{} corresponding to the 2p-1s electronic transition of neutral hydrogen. Historically, observations of the line during flares have been limited, with a particular lack of spectrally resolved observations. While previous studies \citep[e.g.][]{Woods_2004,Milligan_2016} have provided valuable information, there has yet to be a multi-instrument study of a flare using spectrally-resolved \Ly observations. Such a study would provide a deeper understanding of the physical mechanisms driving \Ly spectral variability during flares.

    The \Ly line is optically thick due to the abundance of neutral hydrogen in the chromosphere. In quiet Sun conditions, its wings and core are formed in the mid-chromosphere and lower transition region ($\sim6,000-40,000\;\textrm{K}$), respectively \citep{Vernazza_1973,Vernazza_1981,Fontenla_1991}. In these quiescent conditions, the varied formation heights of different parts of the line wings have been suggested as a diagnostic of density and temperature across the chromosphere \citep{RousselDupre_1983}. However, due to long thermalisation lengths, each wavelength of the \Ly line samples a range of temperatures, limiting it to a secondary diagnostic. These formation conditions undergo drastic changes during flares, as large amounts of energy are deposited into the chromosphere. Flare modelling by \citet{Hong_2019} demonstrated that the formation heights of the \Ly line profile vary significantly between different models. In most flare models driven by nonthermal heating, the \Ly line wings and core form at distinct heights within the chromosphere, both of which are lower than their formation heights under quiescent conditions. This behaviour was attributed to decreasing opacity to wing photons as neutral hydrogen above is excited, with the conditions above the core remaining similar to preflare conditions. However, models with relatively more low-energy nonthermal electrons, as well as a model driven by thermal conduction, showed the wings and core forming at the same height, in the middle chromosphere. This was due to a condensation layer, which lowered the formation region, while the material above was transparent due to significant ionisation. These changes to the line's formation should be considered when interpreting spectrally resolved \Ly flare observations.

    Observations show that different processes during flares can drive enhancement of \Ly in both the chromosphere and corona. In the chromosphere, \Ly enhancement may result from the thermalisation of flare-accelerated electrons with ambient material, where Coulomb collisions excite neutral hydrogen, leading to increased \Ly emission via radiative de-excitation \citep{Kurokawa_1988}.  However, observations from the \Ly channel of the Extreme Ultraviolet Imager's High-Resolution Imager on Solar Orbiter \citep[SolO/EUI HRILya;][]{Muller_2020,Rochus_2020} reveal a more elaborate scenario. \citet{Li_2022} analysed a C-class flare, finding \Ly emission to originate from both flare ribbons and loops. The study demonstrated a strong temporal correlation between \Ly and SXR emission, the primary source of \Ly emission being flare ribbons. Due to this, the authors suggested the \Ly enhancement was predominantly driven by the conduction of energy from coronal flare loops to the chromosphere. Additionally, Hard X-ray (HXR) emission driven by the injection of nonthermal energy was cospatial with these \Ly ribbons, indicating that this injection further contributed to \Ly enhancement. Additional \Ly emission was observed in flare loops, likely driven by the radiative cooling of hot plasma in these structures. \Ly imaging observations of an M-class flare and filament eruption from the Transition Region and Coronal Explorer \citep[TRACE;][]{Handy_1999b} were reported by \citet{RubioDaCosta_2009}. Their analysis demonstrated that \Ly emission originated primarily from flare footpoints and was cospatial with HXR sources, leading the authors to suggest these \Ly enhancements were driven by nonthermal heating of the chromosphere. Further evidence of \Ly flare enhancement associated with filament-eruptions was provided by \citet{Wauters_2022}, who analysed an M6.7 flare with photometric \Ly observations from the Large Yield Radiometer (LYRA) on the Project for On-Board Autonomy 2 \citep[PROBA-2;][]{Dominique_2013}. Their studies revealed emission from a bright filament-eruption captured in $1600\;\textrm{Å}$ images from the Solar Dynamics Observatory's Atmospheric Imaging Assembly \citep[SDO/AIA;][]{Pesnell_2012,Lemen_2012} that correlated with gradual phase \Ly enhancements, suggesting the filament-eruption may have contributed significantly to \Ly enhancement in this event.

    The spectral properties of \Ly during flares remain insufficiently studied, resulting in limited information on how spectral variability (e.g. line asymmetry, variation in enhancements of different regions of the line) varies with changes in flare heating. \citet{Woods_2004} studied an X17 flare on $28$ October 2010 using calibration scans of the \Ly line obtained from the second Solar-Stellar Irradiance Comparison Experiment on board the Solar Radiation and Climate Experiment \citep[SORCE/SOLSTICE;][]{Rottman_2005,Mcclintock_2005}. During this event, the wings of the line were strongly enhanced, by a factor of two, while the line core showed a relatively weaker $20\%$ enhancement. Additionally, a blue asymmetry was observed in the line wings, which the authors did not attribute to a specific mechanism but may have been related to an associated coronal mass ejection. Intriguingly, the blue wing began to enhance prior to the red wing in the flare impulsive phase. It remains unclear whether the observed asymmetry and strong wing enhancement observed during this flare was due to its extreme class, or if similar behaviour is observed in flares of smaller class. To date, comparatively weaker flares have yet to be studied using SORCE/SOLSTICE observations, underscoring the need for such analyses to establish the typical spectral behaviour of \Ly for flares of varying magnitude.

    Flaring \Ly spectra observed using the Naval Research Laboratory (NRL) SO82B spectrograph on Skylab's Apollo Telescope Mount \citep[Skylab/ATM;][]{Bartoe_1977} were studied by \citet{Canfield_1978}. Their analysis revealed a red enhancement asymmetry in the line wings (between $5\;\textrm{Å}<\Delta\lambda_0<12\;\textrm{Å}$) during a flare; although, this asymmetry was not statistically significant. Further observations of two additional flares by Skylab/ATM are presented in \citet{Canfield_1980}. In one flare, a red asymmetry was seen in the line wings (between $0.5\;\textrm{Å}<\Delta\lambda_0<1\;\textrm{Å}$) during the rise phase, transitioning into a blue asymmetry at flare peak. The second flare exhibited a red asymmetry between the two peaks of the centrally reversed \Ly line core, with the red peak culminating at $30-40\%$ brighter than the blue peak. Flaring \Ly line profiles taken by the Laboratoire de Physique Stellaire et Planétaire instrument \citep[LPSP;][]{Bonnet_1978} on the Orbiting Solar Observatory 8 (OSO-8) satellite were presented by \citet{Lemaire_1984}. These profiles suggest a red asymmetry between the two peaks on either side of the line's central reversal and a blue asymmetry further out in the line wings. However, the authors provide limited discussion of these features. \citet{Brekke_1996} observed an X3 flare using the Upper Atmosphere Research Satellite's Solar-Stellar Irradiance Comparison Experiment \citep[UARS/SOLSTICE;][]{Rottman_1993}. Their analysis revealed a blue asymmetry in the \Ly line wings, with a $6\%$ integrated enhancement of the line being measured and the largest relative enhancement occurring in the line wings. These studies highlight the diverse spectral variability in the \Ly during different flares. Further study is necessary to understand how variations in underlying flare mechanisms drive the observed \comment{}changes in \Ly spectral variability\commentend{}.
        
    Although spectrally resolved \Ly flare observations remain scarce, several recent studies have focused on wavelength-integrated observations to study \Ly flare variability. \citet{Milligan_2016} analysed enhancements in the \Ly line during a flare observed with the Extreme Ultraviolet Variability Experiment's Multi EUV Grating Spectrograph Photometer on the Solar Dynamics Observatory \citep[SDO/EVE MEGS-P;][]{Woods_2012}. Their findings revealed good agreement between \Ly and Soft X-ray (SXR) emission. Conversely, a flare observed with SORCE/SOLSTICE exhibited impulsive \Ly emission consistent with the Neupert effect, attributable to differences in instrumental response \citep{Neupert_1968}. Expanding on these findings, \citet{Milligan_2020} conducted a statistical study of \Ly emission during 477 M- and X-class flares observed with the Extreme Ultraviolet Sensor's E channel onboard the Geostationary Operational Environmental Satellites (GOES/EUVS-E) during solar cycle 24. For $95\%$ of these flares, \Ly enhancements remained below $10\%$, with peak emission typically occurring during the impulsive phase. \citet{Milligan_2021} further extended the study to include B- and C-class flares observed by GOES/EUVS-E. Average \Ly enhancements were $0.1-0.3\%$, substantially weaker than the $1-4\%$ enhancements typically observed during M- and X-class flares. Notably, a C6.6 flare exhibited a $7\%$ enhancement, that was attributed to a failed filament-eruption. \citet{Greatorex_2023} further investigated \Ly variability by comparing \Ly and HXR emission in three flares of equivalent GOES class observed with GOES/EUVS-E and the Reuven Ramaty High Energy Solar Spectroscopic Imager \citep[RHESSI;][]{Lin_2002}. Their studies revealed event-dependent variations in \Ly enhancement, with flares that exhibited larger nonthermal electron spectral indices producing greater \Ly enhancements, highlighting the importance of these nonthermal electron properties in understanding \Ly flare variability. These findings have provided significant insights into \Ly flare emission in the absence of spectrally resolved \Ly observations.
    
    This paper presents spectrally-resolved \Ly observations from SORCE/SOLSTICE for two M-class flares. The potential physical mechanisms responsible for flare enhancements across the \Ly line profile are evaluated using a multi-instrument approach. Key observations include HXR count rates from RHESSI, imaging from SDO/AIA, and observations from the Extreme Ultraviolet Imager \citep[EUVI;][]{Wuesler_2004} of the Sun-Earth Connection Coronal and Heliospheric Investigation \citep[SECCHI;][]{Howard_2008}, onboard the two Solar TErrestrial RElations Observatory \citep[STEREO;][]{Kaiser_2008} satellites. To validate the SORCE/SOLSTICE \Ly observations, comparisons were made with photometric \Ly observations from GOES/EUVS-E, an established \Ly instrument. Section \ref{S-Observations} outlines the instruments used in this study, along with the methodology and analysis applied. The results of this analysis are presented in Section \ref{s-results}. Finally, Section \ref{S- Conclusions} discusses the findings and their implications, along with an overview of the study's contribution to understanding \Ly flare variability, with insights for future and contemporary \Ly instrumentation.    
\section{Observations, Methodology \& Analysis}
    \label{S-Observations}
    This study examines two flares: an M8.3 flare on $12$ February 2010 (\flrone) in NOAA active region 11046 (N47E16) and an M5.3 flare on $4$ July 2012 (\flrtwo) in NOAA active region 11515 (S31W27). The selection of the flares was based on the following criteria:
    
    \begin{itemize}
    \item Coverage of the impulsive phase by SOLSTICE, RHESSI and \mbox{GOES/EUVS-E}
    \item Complete or partial imaging coverage at UV wavelengths 
    \item Close proximity to disk-centre to minimise limb-darkening \citep{Milligan_2020, Milligan_2021}
    \end{itemize}
    
\begin{figure}    
        \centerline{\hspace*{0.015\textwidth}
               \includegraphics[width=.65\textwidth,clip=]{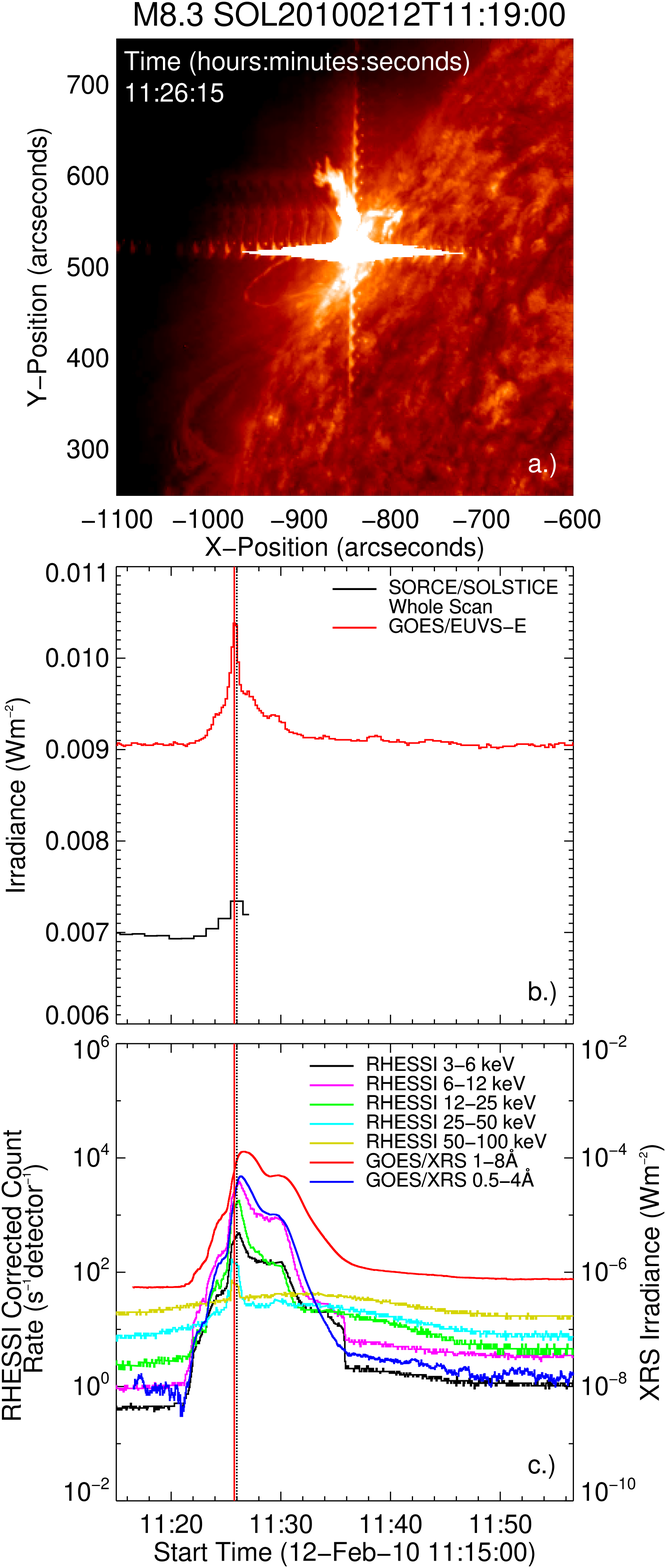}
               \hspace*{-0.0\textwidth}
               \includegraphics[width=.65\textwidth,clip=]{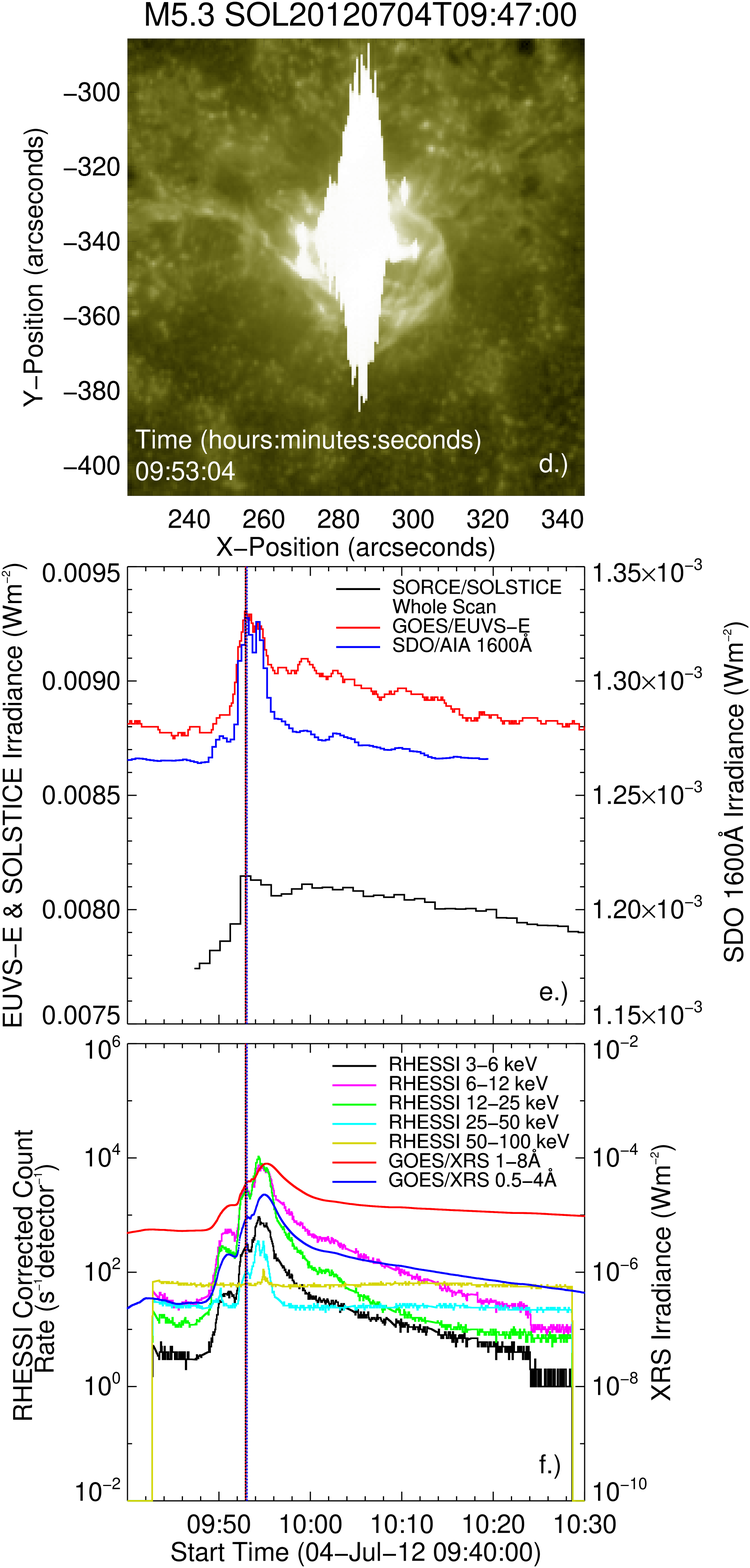}
              }
    \caption{Top panels: Context images from STEREO-A/SECCHI EUVI $304\;\textrm{Å}$ (left) and uncorrected SDO/AIA $1600\;\textrm{Å}$ (right). Middle panels: Integrated \Ly irradiances from SORCE/SOLSTICE and GOES/EUVS-E, with disk-integrated SDO/AIA $1600\;\textrm{Å}$ irradiance for \flrtwos only. Bottom panels: X-ray corrected count rates and irradiance in different energy channels from RHESSI and GOES/XRS, respectively. Vertical lines indicate peaks in UV emission from lightcurves of the same colour.}
    \label{Figure1}
    \end{figure}

    Figure \ref{Figure1} illustrates observations for both events. Panels a.), b.) and c.) display data for \flrone. Specifically, panel a.) shows a He \textsc{ii} $304\;\textrm{Å}$ image of \flrones from STEREO-A/SECCHI EUVI, panel b.) shows lightcurves from both GOES/EUVS-E and SOLSTICE, and panel c.) shows HXR and SXR lightcurves from RHESSI and the X-ray Sensor on GOES (GOES/XRS). Similarly, panels d.), e.) and f.) depict observations for \flrtwo, with panel d.) featuring an SDO/AIA $1600\;\textrm{Å}$ image, panel e.) presenting lightcurves from SORCE/SOLSTICE, GOES/EUVS-E and SDO/AIA $1600\;\textrm{Å}$, and panel f.) showing HXR and SXR lightcurves. 
     Lightcurves were generated from \Ly spectra for distinct wavelength ranges within the \Ly line to explore its spectral variability. Comparisons were made between the timings and magnitudes of enhancement in these SORCE/SOLSTICE lightcurves and in photometric observations from GOES/EUVS-E, providing validation of SORCE/SOLSTICE measurements. Further analysis examined the temporal correlation between SORCE/SOLSTICE enhancements and both HXR and SXR enhancements observed by RHESSI and GOES/XRS, respectively. This allowed for the assessment of whether \Ly spectral variability was primarily driven by impulsive or gradual phase processes. To provide additional context, SORCE/SOLSTICE observations were further compared to imaging observations, enabling the separation of chromospheric and coronal contributions to \Ly enhancement.

   \begin{figure}    
        \centerline{\hspace*{0.015\textwidth}
               \includegraphics[width=0.7\textwidth,clip=]{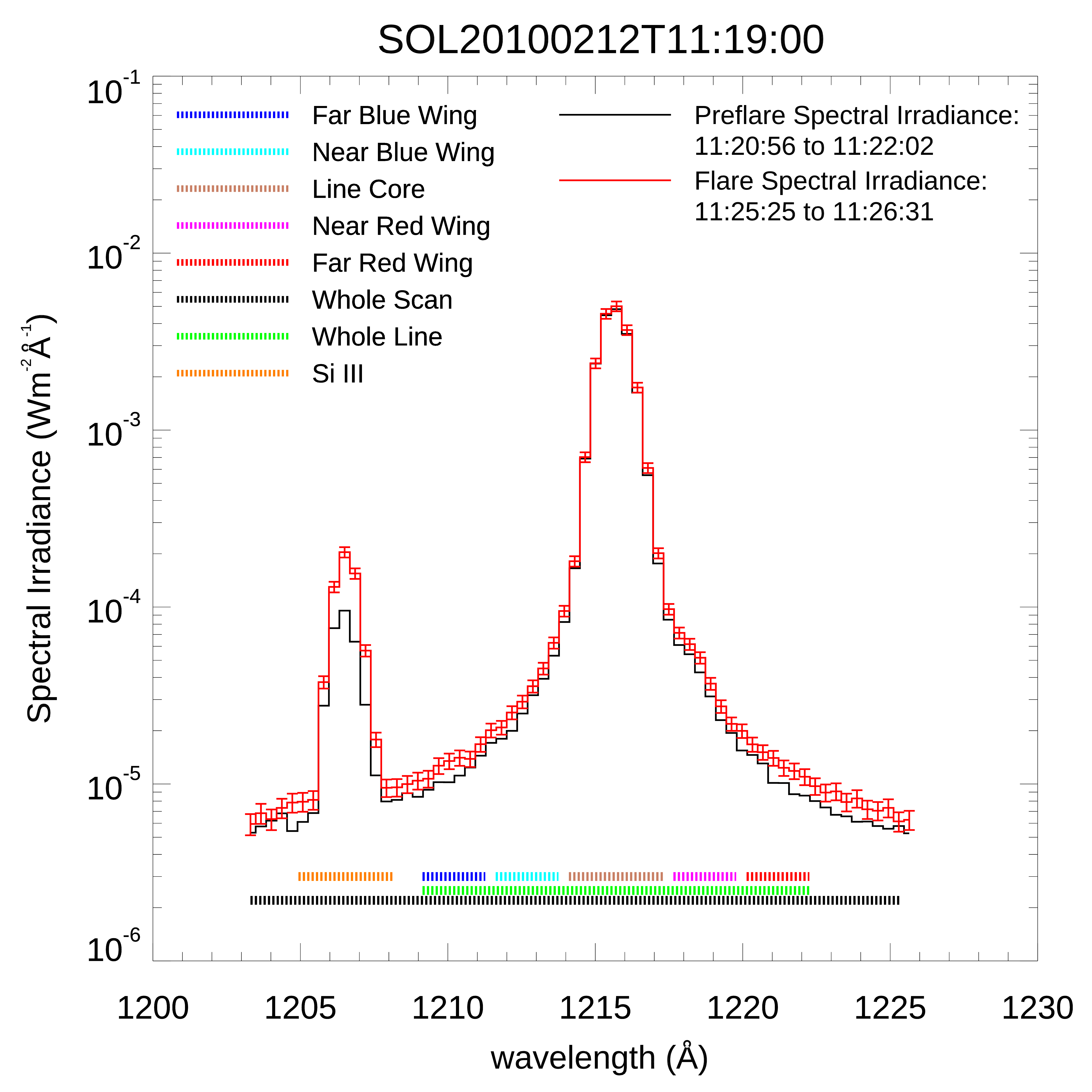}
               \hspace*{-0.03\textwidth}
               \includegraphics[width=0.7\textwidth,clip=]{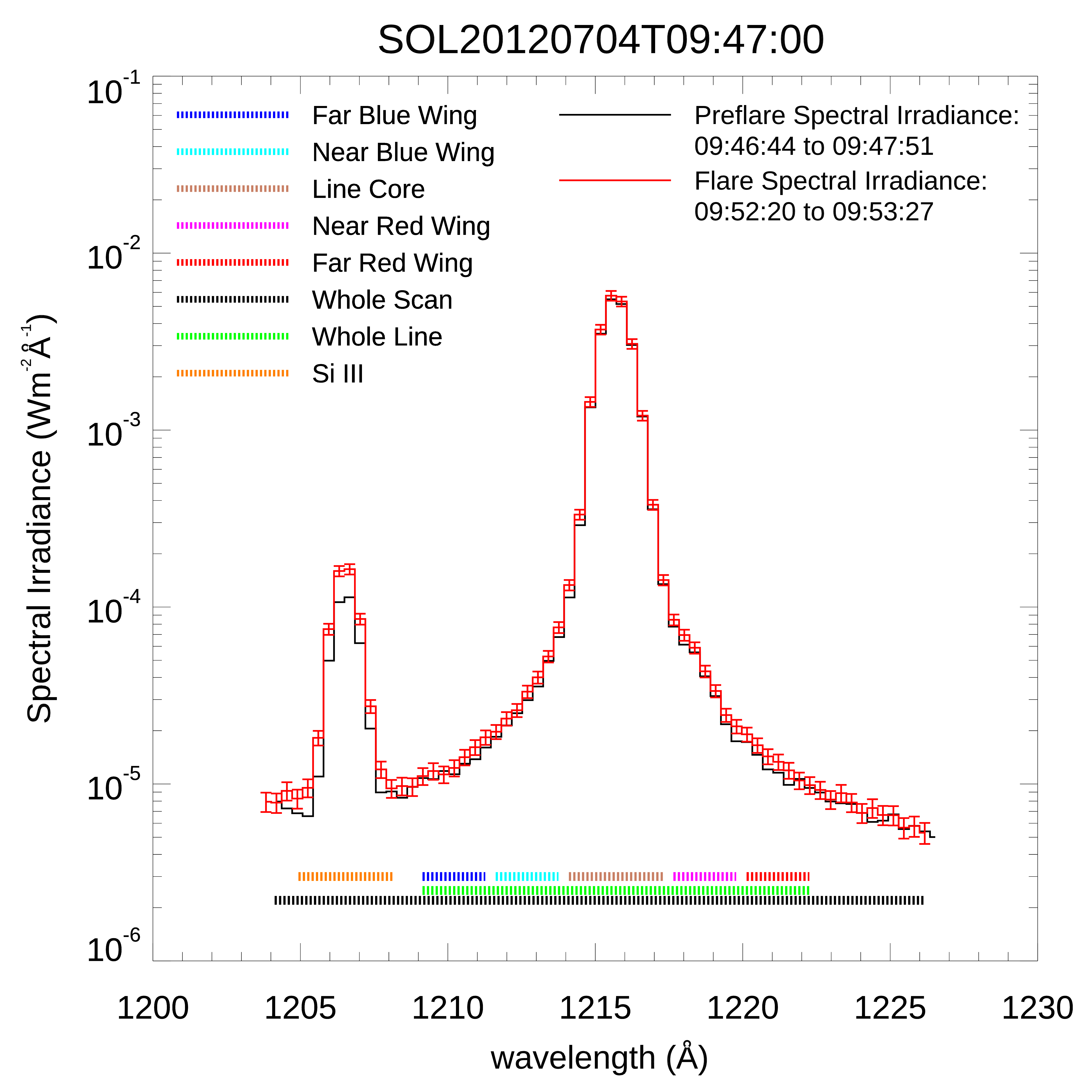}
              }
    \caption{SORCE/SOLSTICE line profiles during \flrones (left) and \flrtwos (right). For each, a raster scan taken at the time of peak integrated \Ly flare enhancement is plotted as a red line with uncertainty plotted as error bars and a preflare raster is plotted as a black line. 
        }
   \label{Figure2}
   \end{figure}
    
    \subsection{SORCE/SOLSTICE}    
    
    SORCE/SOLSTICE performed disk-integrated wavelength calibration scans through the \Ly line once per orbit from 2003 until 2012, with an effective raster cadence of $\sim1$ minute, suitable for the study of spectral behaviour during flares \citep{Woods_2004,Snow_2022}. Recently, these data have been made publicly available via the version 18 release of level 3 SORCE/SOLSTICE data products, which are hosted in the SORCE data repository\footnote{https://lasp.colorado.edu/sorce/data/}. The left and right panels of Figure \ref{Figure2} show SORCE/SOLSTICE \Ly line profiles for \flrones and \flrtwo, respectively. 

    The wavelength calibration scans consist of rastered spectral irradiance measurements taken between $1203\;\textrm{Å}$ and $1227\;\textrm{Å}$ with a wavelength resolution of $0.35\;\textrm{Å}$. Each spectral irradiance measurement typically took $1.05\;\textrm{s}$, and a complete raster of $64$ measurements through the entire wavelength range took around $67\;\textrm{s}$. The duration of a full scan varied but was typically around an hour. A number of these scans coincided with flares, providing some of the only wavelength and flux-calibrated spectrally-resolved observations of \Ly during flares. Further discussion of the SORCE/SOLSTICE wavelength calibration scan observations can be found in \citet{Snow_2022}. While significant degradation has been observed in other \Ly instruments, such as GOES/EUVS-E and PROBA-2/LYRA, no degradation correction has been applied to these scans \citep{Greatorex_2024}. As a result, the irradiance values may contain uncorrected systematic errors. To assess the potential impact of such errors on the timing of flare enhancements, comparisons are made between SORCE/SOLSTICE and GOES/EUVS-E observations. 
    
    To probe the behaviour of different parts of the \Ly line during flares, the SORCE/SOLSTICE data were divided into several spectral bands, as defined in Table~\ref{Table1} and illustrated in Figure \ref{Figure2}. For each band, data points (in Wm$^{-2}$nm$^{-1}$) were integrated to provide a single irradiance value (in Wm$^{-2}$) per raster scan. The $1\sigma$ combined standard uncertainties provided in the SORCE/SOLSTICE dataset, were propagated to provide the uncertainty in the spectrally-integrated irradiance values. 
    
    The Whole Scan band (black shaded bar in Figure~\ref{Figure2}) includes every data point in each raster, covering both the \Ly and Si \textsc{iii} ($\sim1206.5\;\textrm{Å}$) doublet. Also included are the O \textsc{v} line blend at $1218.34\;\textrm{Å}$ in the \Ly near red wing, and the comparatively weaker He \textsc{ii} line doublet ($\sim1215.1\;\textrm{Å}$) blended in the \Ly blue wing \citep[CHIANTI\footnote{https://www.chiantidatabase.org/chianti\_linelist.html} Version 11.0;][]{Dere_1997,Dufresne_2024}. The Whole Line band (bright green shaded bar in Figure \ref{Figure2}) omits Si \textsc{iii} emission and includes data points for only the \Ly line ($1209.3-1221.9\;\textrm{Å}$). The remaining bands were defined by specific data points of each raster, rather than fixed wavelength ranges. This was necessary because SORCE/SOLSTICE data points slowly drift in wavelength between raster scans, which can cause data points to enter or exit a given band, leading to discontinuous `jumps' in irradiance. Such effects do not impact the Whole Scan band, as it encompasses all data points, and have minimal impact on the Whole Line band due to the much greater irradiance near the line centroid relative to the band's edges, making any `jumps' in irradiance negligible. The Line Core band (brown shaded bar in Figure \ref{Figure2}) comprises nine data points, with the central point being the closest to the centroid of the \Ly line for a given raster. The Near Wing bands (magenta and cyan shaded bars in Figure \ref{Figure2}) contain the nearest six points outside of the Line Core, positioned to the red or blue of the central wavelength. The Far Wings (red and blue shaded bars in Figure \ref{Figure2}) include the nearest five points outside of both the Line Core and Near Wings. Finally, the Si \textsc{iii} band (orange shaded bar in Figure \ref{Figure2}) was defined similarly to the Line Core band, including nine data points centred around the Si \textsc{iii} line at $1206\;\textrm{Å}$.

    For \flrones, background values for each band were determined by averaging five band integrated measurements before the GOES start time (from 11:14:40 UT to 11:18:09 UT). For \flrtwo, since the SORCE/SOLSTICE scan began slightly after the flare start time, the backgrounds were instead defined as the integrated irradiance measured during the first raster of the SORCE/SOLSTICE observation for each band (09:47:18 UT). This may lead to a slight underestimation of flare enhancements, which were quantified as the percentage increase in irradiance above their respective background value. 
    
    The timings and magnitudes of peak flare enhancements were determined for each band, enabling the study of spectral variability in the \Ly line during the two flares. Asymmetry in the Near and Far Wing bands was quantified by calculating the difference in percentage enhancement between the Near Red Wing and Near Blue Wing bands, as well as between the Far Red and Far Blue Wings bands. It should be noted that rastering may introduce instrumentally driven asymmetries, as variation in flare enhancement in the wings may have occurred between the times each wing was scanned.
        
    \begin{table}
        \caption{Wavelength ranges for the different SORCE/SOLSTICE bands defined for this study.}
        \begin{tabular}{ccclc}     
          \hline                   
            Band & Wavelength Range (Å) & Width (Å) & $\Delta t$ (s) & Data Points \\ 
            \hline
            Whole Scan & $1203.0-1227.0$ & $24$ & $67$ & $64$\\ 
            Whole Line & $1209.3 - 1221.9$ & $12.6$ & $35$ & $\sim36$\\
            Si \textsc{iii} & $\sim1205.0 - 1208.0$ & $3$ & $9.45$ & $9$\\
            Line Core & $\sim1214.1 - 1217.1 $ & $3$ & $9.45$ & 9\\
            Near Red Wing & $\sim1217.5 - 1219.5$  & $2$ & $7.35$ & $6$\\
            Near Blue Wing & $\sim1211.7 - 1213.7$ & $2$ & $7.35$ & $6$\\
            Far Red Wing & $\sim1219.9 - 1221.9$ & $2$ & $7.35$ & $6$\\
            Far Blue Wing & $\sim1209.3 - 1211.3$ & $2$ & $7.35$ & $6$\\
            \hline
        \end{tabular}
        \label{Table1}
    \end{table}      
    
    \subsection{GOES/EUVS-E \& GOES/XRS}  
    
    The E channel of EUVS on the GOES-N satellites (GOES/EUVS-E) provided broadband photometry at a cadence of $10.24\;\textrm{s}$, covering a wavelength range between $1180\;\textrm{Å}$ and $1270\;\textrm{Å}$. This range includes the strong \Ly and Si \textsc{iii} lines, as well as several weaker lines. GOES/EUVS-E has been a valuable resource for \Ly observations during flares, with statistical studies of the data made by \citet{Milligan_2020} and \citet{Milligan_2021}. EUVS-E conducted observations between 2006 and 2016, while from 2017 onward, observations have been provided by the B channel of EUVS on GOES-R \citep[GOES/EUVS-B;][]{Eparvier_2009}. For \flrone, GOES-14 observations were utilised, and for \flrtwo, GOES-15 observations were employed. This leads to potential differences in instrumental performance during each flare. Version 5, level 2 data for both flares was downloaded from the NOAA archive\footnote{https://www.ncei.noaa.gov/data/goes-space-environment-monitor}. These data are scaled to SORCE/SOLSTICE low-cadence observations, providing irradiance calibration, with an additional scaling factor applied to correct instrumental degradation. For each flare, the background was defined as the mean irradiance in a period of 20 minutes before the GOES flare start time, and the standard deviation of irradiance in this period was used to quantify uncertainty. Relative irradiance during each flare was calculated by dividing the measured irradiance by the background value. Flare enhancements and their timings were then determined, allowing for comparison with the timing and magnitude of enhancements measured by SOLSTICE, providing validation. Since the response function of GOES/EUVS-E weights the \Ly and Si \textsc{iii} line similarly, comparing GOES/EUVS-E enhancements to those observed in SORCE/SOLSTICE offers insight into the relative significance of each line to the wavelength-integrated flare enhancements measured by GOES/EUVS-E during \flrones and \flrtwos \citep{Machol_2014}. 
    
    The GOES series of satellites also carry the GOES/XRS instrument, which provides broadband observations of X-rays at a cadence of $2\;\textrm{s}$ between $1-8\;\textrm{Å}$ and $0.5-4\;\textrm{Å}$ via its long and short channels, respectively. Level 2 GOES/XRS data was downloaded from the NOAA archive, and the timings of enhancements in each GOES/XRS channel were compared to \Ly enhancements observed by SOLSTICE. Given that SXR emission characterises the gradual phase of flares, this comparison serves to determine whether \Ly spectral enhancements are driven by gradual phase processes. This analysis is further supported by comparisons to RHESSI observations, which are detailed in Section \ref{RHESSI}. 
        
    \subsection{RHESSI}
    \label{RHESSI}
    During flares, magnetic reconnection drives the acceleration of electrons to high energies ($\sim10-100\;\textrm{keV}$). These electrons undergo thick target collisions in the chromosphere, leading to HXR bremsstrahlung emission \citep{Brown_1971,Kontar_2011}. RHESSI provided photometrically calibrated spectroscopic and imaging observations of X-rays and $\gamma$-rays between $3$ keV and $17$ MeV, with an energy resolution of $\sim1-10$ 
    keV, attaining higher resolution at lower energies. To mitigate the effects of detector degradation, RHESSI periodically underwent anneals. Since an anneal was performed closer to the time of \flrtwo, detector sensitivity may have been relatively reduced during \flrone. Level 1 RHESSI data for both flares was retrieved from the RHESSI data archive using the observing summary object in SolarSoft. These data include binned X-ray count rates in a number of different channels (including $6-12, 25-50, 50-100\;\textrm{keV}$) with a cadence of $4\;\textrm{s}$. Comparisons between enhancements in the HXR $25-50\;\textrm{keV}$ and SXR $6-12\;\textrm{keV}$ bands with SORCE/SOLSTICE observations provided insights on whether \Ly enhancements were driven by impulsive nonthermal heating or by thermal gradual phase processes. 
         
    \subsection{STEREO/SECCHI EUVI}
    
    The two STEREO/SECCHI EUVI instruments provide full disk imaging of the Sun in four wavelength bands ($171\;\textrm{Å}$, $195\;\textrm{Å}$, $284\;\textrm{Å}$,$304\;\textrm{Å}$) with a cadence between $2.5$ and $10$ minutes. Together, STEREO-A and STEREO-B provided a stereoscopic view of the Sun. He \textsc{ii} $304\;\textrm{Å}$ emission, which is a potential proxy for hydrogen \Ly emission in the quiet Sun \citep{Auchere_2005, Gordino_2022}, is extended to the flaring Sun in this study. STEREO/SECCHI EUVI data were downloaded from the Stereo Science Centre and processed to level 1 using SolarSoft's \texttt{secchi\_prep} routine. The $304\;\textrm{Å}$ images were visually inspected to determine whether \Ly flare enhancement was likely confined to flare ribbons or if there was a contribution from coronal flare loops. This analysis was performed for \flrone, as it occurred prior to the availability of higher resolution, higher cadence imaging from SDO/AIA. These SDO/AIA observations enable a more systematic separation of coronal and chromospheric contributions. 
    
    \subsection{SDO/AIA}
    
    SDO/AIA provides high spatial resolution ($1^{\prime\prime}$) imaging of the entire solar disk in nine different wavelength bands, with a cadence of $12\;\textrm{s}$ for its extreme ultraviolet (EUV) channels and $24\;\textrm{s}$ for its UV channels. The UV $1600\;\textrm{Å}$ channel, which is dominated by chromospheric emission, was used to provide context imaging for similarly chromospheric \Ly emission during \flrtwos \citep{Simoes_2019}. The choice of $1600\;\textrm{Å}$ images over similarly available He \textsc{ii} $304\;\textrm{Å}$ images from SDO/AIA was based on the significantly lower saturation observed in the $1600\;\textrm{Å}$ images during the flare, allowing for clearer separation of spatial features. SDO/AIA data was downloaded from the JSOC data archive and processed to level 1.5 using the \texttt{aia\_prep} routine in SolarSoft. Degradation to the UV channels of SDO/AIA can be corrected using SolarSoft's \texttt{aia\_get\_response} routine. This correction is applied in the lightcurves seen in panel e.) of Figure \ref{Figure1} but is not applied for the analysis of desaturated images. 

    To correct for strong saturation in the $1600\textrm{Å}$ images, the methods of \citet{Kazachenko_2017} were applied. Saturated pixels ($>5000\;\textrm{DNs}^{-1}$) and pixels within a two-pixel radius in $x$ direction and a 10-pixel radius in $y$ direction from the saturated pixels were selected. The intensity values of these pixels during the saturated period were replaced using values determined from a linear interpolation between the pre-saturated and the post-saturated intensities for the same pixel. Different flare features were identified in the desaturated images by visual inspection, with sections within the images being defined for each feature. The pixel intensities of each section were then summed to generate lightcurves. The timing and magnitude of enhancements in each of these features were then compared to  SORCE/SOLSTICE spectral \Ly enhancements, offering insight into which features, and hence which physical processes, contributed to the spectral \Ly emission observed during \flrtwo. 
        
\section{Results}
    \label{s-results}
    \subsection{SOL2010-02-12T11:19}
        \label{S-Flare_One}

            \begin{figure}
            \centerline{\hspace*{0.015\textwidth}
                       \includegraphics[width=0.7\textwidth,clip=]{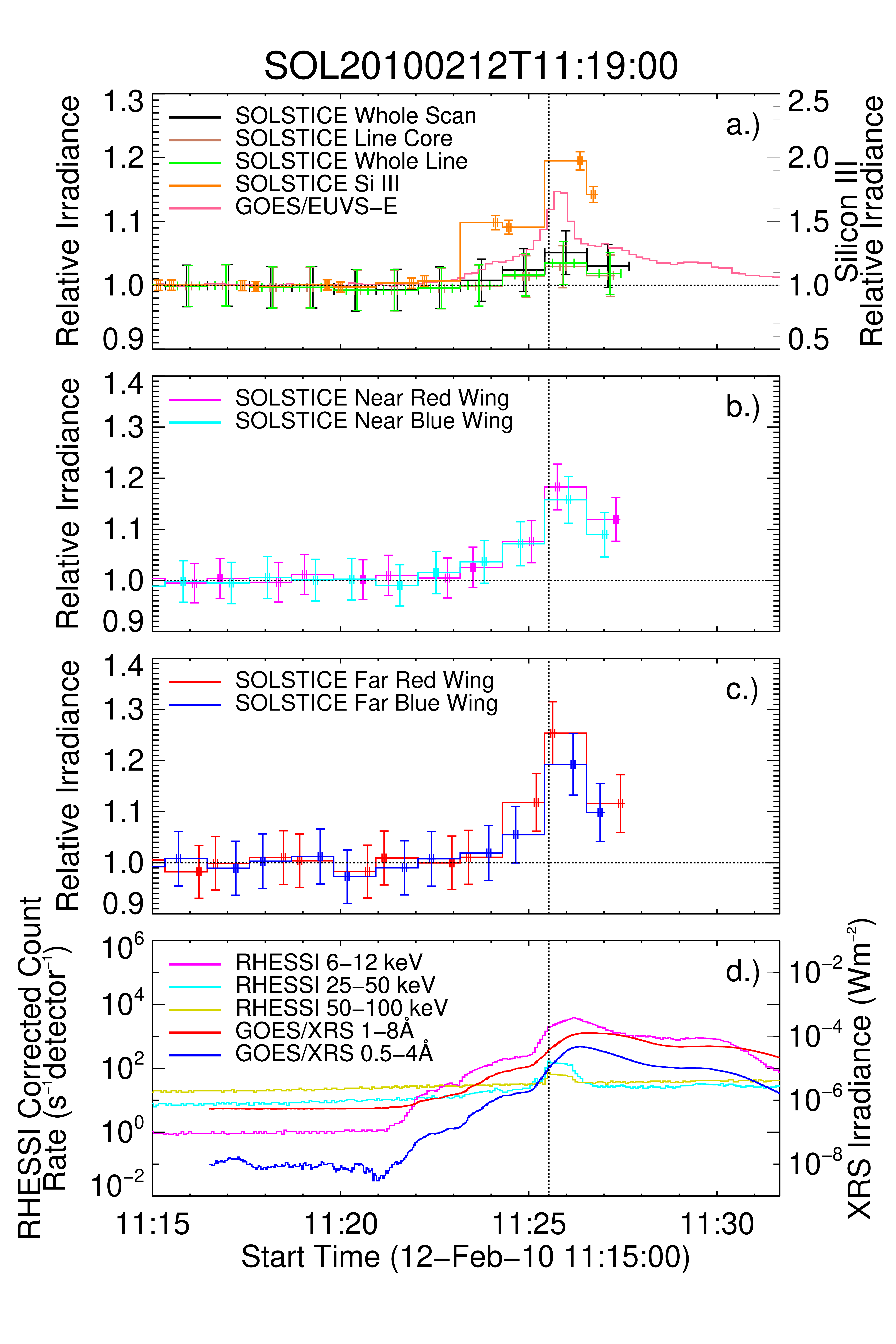} 
                      }
            \caption{Relative irradiance to the background in each SORCE/SOLSTICE band and GOES/EUVS-E, with RHESSI detector averaged corrected count rates. a.), b.) and c.) show relative enhancements of each SORCE/SOLSTICE band during \flrone. d.) shows HXR count rates from RHESSI in its $6-12\;\textrm{keV}$, $25-50\;\textrm{keV}$ and $50-100\;\textrm{keV}$ channels along with irradiances from GOES/XRS long and short channels. A black dashed line indicates the peak of a HXR burst in the $25-50\;\textrm{keV}$ channel. 
                }
            \label{Figure3}
            \end{figure}
            
            Figure \ref{Figure3} illustrates the time evolution of relative irradiance in each SORCE/SOLSTICE band for \flrone, with peak flare enhancements listed in Table \ref{Table2}. During this event, the strongest absolute enhancement in the \Ly line was observed in the Line Core band, reaching $1.91\times10^{-4}\;\textrm{Wm}^{-2}$, as seen in panel a.) of Figure \ref{Figure3}. While it was strong in absolute terms, this was only a modest relative enhancement. Absolute enhancements the Near Wing bands were over an order of magnitude lower, with the Near Red Wing being enhanced by $1.49\times10^{-5}\;\textrm{Wm}^{-2}$ and the Near Blue Wing being enhanced by $1.06\times10^{-5}\;\textrm{Wm}^{-2}$. Panel b.) of Figure \ref{Figure3} reveals a $2.5\%$ greater peak relative enhancement in the Near Red Wing compared to the Near Blue Wing, exhibiting a slight red enhancement asymmetry, though this difference falls within uncertainty. The observed asymmetry in the Near Wings may also be influenced by the O \textsc{v} ($1218.34\;\textrm{Å}$) line blend. The Far Wing bands had the smallest absolute enhancements in the \Ly line, with the Far Red and Far Blue Wings being enhanced by $5.82\times10^{-6}\;\textrm{Wm}^{-2}$ and $4.67\times10^{-6}\;\textrm{Wm}^{-2}$, respectively. Panel c.) of Figure \ref{Figure3} shows a red enhancement asymmetry in the Far Wings, with the Far Red Wing exhibiting an enhancement $6.1\%$ greater than the Far Blue Wing, slightly exceeding uncertainty. The Si \textsc{iii} line showed strong absolute enhancement of $1.10\times10^{-4}\;\textrm{Wm}^{-2}$, with a substantially higher relative enhancement ($97.5\%$) compared to the \Ly bands.
            
            As shown in panels a.), b.) and c.) of Figure \ref{Figure3}, the relative enhancement of each Wing band was greater than the Line Core band, consistent with results from \citet{Brekke_1996} and \citet{Woods_2004}. Panel d.) of Figure \ref{Figure3} presents HXR observations from RHESSI in the $6-12$, $25-50$ and $50-100\;\textrm{keV}$ channels, alongside both channels of GOES/XRS. Good agreement is seen between the timing of a HXR burst and \Ly emission across the line profile. The $25-50\;\textrm{keV}$ channel peaked at 11:25:32 UT, aligning with peak enhancements in each SORCE/SOLSTICE band, which all peaked within the same raster (median time 11:25:59 UT). GOES/EUVS-E recorded peak broadband \Ly emission at 11:25:45 UT, closely coinciding with the HXR peak. Of the SORCE/SOLSTICE bands, the peak HXR emission occurred closest in time to the Far Red Wing band, which peaked at 11:25:38 UT. Imaging from STEREO/SECCHI EUVI $304\;\textrm{Å}$ (panel a.) of Figure \ref{Figure1}), though heavily saturated, reveals enhancements in both the chromosphere and corona at 11:26:15 UT.
            
            \begin{table}
            \caption{Flare enhancements, timings and irradiances along with their uncertainties quoted at $1\sigma$ are presented here for GOES/EUVS-E and each SORCE/SOLSTICE band during \flrone.}
            \begin{tabular}{ccccc}     
            \toprule                   
            \flrone & \thead{Peak Relative \\ Flare Enhancement} & \thead{Peak Irradiance \\ (Wm$^{-2}$)} & \thead{Peak Excess \\ Irradiance (Wm$^{-2}$)} & Median Time (UT) \\   
            \toprule
            GOES/EUVS-E & $14.7\pm0.3\%$ & $10.38\pm0.03\times10^{-3}$ & $1.33\pm0.03\times10^{-3}$ & 11:25:45 \\
            \toprule
            SORCE/SOLSTICE& & & & \\
            \hline
            Whole Scan & $5.1\pm3.3\%$ & $7.34\pm0.24\times10^{-3}$ & $3.56\pm2.40\times10^{-4}$ & 11:25:59 \\
            Si \textsc{iii} & $97.5\pm3.6\%$ & $2.22\pm0.08\times10^{-4}$ & $1.10\pm0.08\times10^{-4}$ & 11:26:21 \\
            Whole Line & $3.5\pm3.2\%$ & $7.07\pm0.23\times10^{-3}$ & $2.38\pm2.28\times10^{-4}$ & 11:25:55 \\
            Line Core & $2.9\pm3.2\%$ & $6.76\pm0.22\times10^{-3}$ & $1.91\pm2.16\times10^{-4}$ & 11:25:54 \\
            Near Red Wing & $18.3\pm3.8\%$ & $9.61\pm0.36\times10^{-5}$ & $1.49\pm3.64\times10^{-5}$& 11:25:45 \\
            Near Blue Wing & $15.8\pm4.0\%$ & $7.76\pm0.31\times10^{-5}$ & $1.06\pm3.07\times10^{-5}$ & 11:26:03 \\
            Far Red Wing & $25.4\pm4.9\%$ & $2.87\pm0.14\times10^{-5}$ & $5.82\pm1.41\times10^{-6}$ & 11:25:38 \\
            Far Blue Wing & $19.3\pm5.0\%$ & $2.89\pm0.15\times10^{-5}$ & $4.67\pm1.46\times10^{-6}$ & 11:26:11 \\
            \toprule
            \end{tabular}
            \label{Table2}
            \end{table} 

            Panel a.) of Figure \ref{Figure3} compares the measured \Ly emission from SORCE/SOLSTICE Whole Scan and GOES/EUVS-E for \flrone. GOES/EUVS-E measured peak flare enhancement at 11:25:45 UT, which was within the raster at which SORCE/SOLSTICE Whole Scan had its peak enhancement at 11:25:59 UT. However, the peak relative enhancements were markedly different, with SORCE/SOLSTICE measuring $5.1\%$ and GOES/EUVS-E measuring $14.7\%$. Panel a.) also illustrates that SORCE/SOLSTICE observed much greater relative flare enhancement of the Si \textsc{iii} line at $97.5\%$ than the \Ly Whole Line band at $3.5\%$. Despite the \Ly line being much stronger than the Si \textsc{iii} line in quiescent conditions, the Si \textsc{iii} band had a comparable peak excess irradiance of $1.08\times10^{-4}$ Wm$^{-2}$ to the $2.38\times10^{-4}$ Wm$^{-2}$ of the Whole Line band for \Lyns. This suggests that the Si \textsc{iii} line may contribute significantly to the flare excess measured by GOES/EUVS-E as both lines have a similar weighting in the instrument's response function \citep{Machol_2014}.
             
    \subsection{SOL2012-07-04T09:47}
        \label{S-Flare_Two}    
            
            \begin{figure}    
            \centerline{\hspace*{0.015\textwidth}
                       \includegraphics[width=0.7\textwidth,clip=]{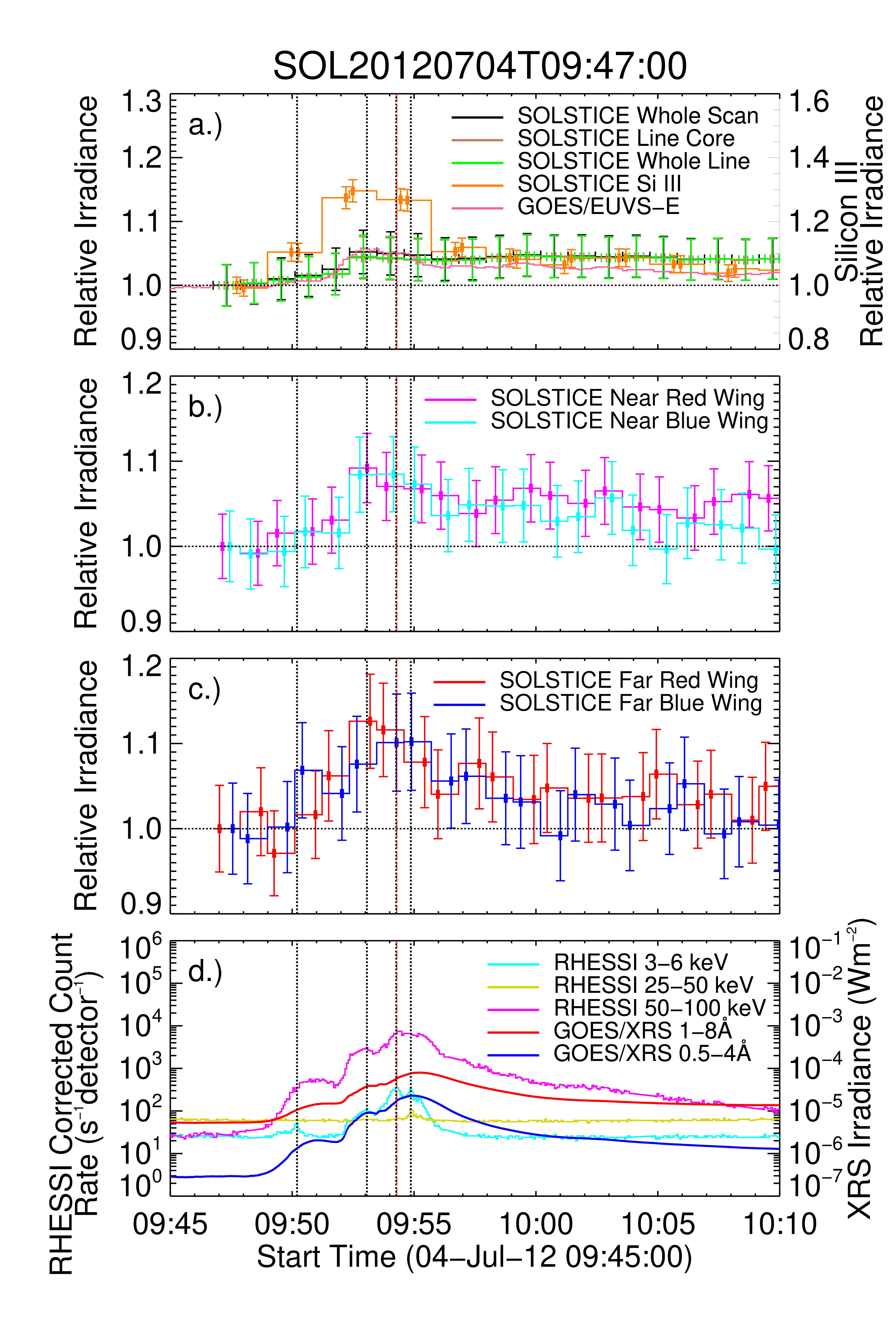} 
                      }
            \caption{Relative irradiance to the background in each SORCE/SOLSTICE band and GOES/EUVS-E, with RHESSI detector averaged corrected count rates. a.), b.) and c.) show relative enhancements of each SORCE/SOLSTICE band during \flrtwo. d.) shows HXR count rates from RHESSI in its $6-12\;\textrm{keV}$, $25-50\;\textrm{keV}$ and $50-100\;\textrm{keV}$ channels along with irradiances from GOES/XRS long and short channels. Black dashed lines indicate peaks of HXR bursts in the $25-50\;\textrm{keV}$ channel. The red dashed line indicates a peak in emission from a bright erupting filament seen in SDO/AIA $1600\;\textrm{Å}$ images. 
                }
            \label{figure4}
            \end{figure}
            
             The time evolution of relative irradiance in each SORCE/SOLSTICE band for \flrtwos is shown in Figure \ref{figure4}, with peak flare enhancements listed in Table \ref{table3}. As with \flrone, the \Ly line exhibited its strongest absolute enhancement in the Line Core band during \flrtwo, reaching $3.34\times10^{-4}\;\textrm{Wm}^{-2}$. The Near Wing bands showed smaller absolute enhancements, with the Near Red Wing enhancement peaking at $9.38\times10^{-6}\;\textrm{Wm}^{-2}$ and the Near Blue Wing enhancement peaking at $5.41\times10^{-6}\;\textrm{Wm}^{-2}$. Panel b.) of Figure \ref{figure4} reveals that the Near Red Wing band's peak enhancement was $0.8\%$ more than the Near Blue Wing, though this was within uncertainty. The asymmetry may have also been contributed to by the O \textsc{v} line blend. This red enhancement asymmetry in the Near Wing bands later transitioned to blue as the Near Blue Wing peaked, exhibiting a $1.4\%$ greater enhancement than the Near Red wing during the same raster, though this difference also fell within uncertainty. 
             
             The Far Wing bands had further diminished absolute enhancements, with the Far Red Wing enhancement peaking at $3.04\times10^{-6}\;\textrm{Wm}^{-2}$ and the Far Blue Wing enhancement peaking at $2.73\times10^{-6}\;\textrm{Wm}^{-2}$. As illustrated in panel c.) of Figure \ref{figure4}, the Far Red Wing enhancement exceeded that of the Far Blue Wing by $5.0\%$ at its peak, a difference greater than the uncertainty of the Far Red Wing, but less than that of the Far Blue Wing. Later, this enhancement asymmetry also changed to blue as the Far Blue Wing peaked $2.4\%$ higher than the Far Red Wing during the same raster, though this was within uncertainty.
             
            \begin{table}[ht]
            
            \caption{Flare enhancements, timings and irradiances along with their uncertainties quoted at $1\sigma$ are presented here for GOES/EUVS-E and each SORCE/SOLSTICE band during \flrtwo. 
            }
            \begin{tabular}{cccccc}     
            \toprule
            \flrtwo & \thead{Peak Relative \\ Flare Enhancement} & \thead{Peak Irradiance \\ (Wm$^{-2}$)} & \thead{Peak Excess \\ Irradiance (Wm$^{-2}$)} & Median Time (UT) \\   
            
            \toprule
            GOES/EUVS-E & $5.8\pm0.2\%$ & $9.30\pm0.02\times10^{-3}$ & $5.12\pm0.16\times10^{-4}$ & 09:52:55 \\
            \toprule
            SORCE/SOLSTICE  & & & & \\
            \hline
            Whole Scan & $5.2\pm3.2\%$ & $8.15\pm0.26\times10^{-3}$ & $4.04\pm2.63\times10^{-4}$ & 09:52:55 \\
            Si \textsc{iii} & $44.4\pm3.6\%$ & $1.99\pm0.07\times10^{-4}$ & $6.11\pm0.72\times10^{-5}$ & 09:52:29 \\
            Whole Line & $4.6\pm3.2\%$ & $7.90\pm0.25\times10^{-3}$ & $3.51\pm2.50\times10^{-4}$ & 09:59:39 \\
            Line Core & $4.6\pm3.1\%$ & $7.59\pm0.24\times10^{-3}$ & $3.34\pm2.37\times10^{-4}$ & 09:59:39 \\
            Near Red Wing & $9.2\pm3.7\%$ & $1.11\pm0.04\times10^{-4}$ & $9.38\pm4.15\times10^{-6}$ & 09:53:04 \\
            Near Blue Wing & $8.5\pm4.1\%$ & $6.92\pm0.28\times10^{-5}$ & $5.41\pm2.82\times10^{-6}$ & 09:54:10 \\
            Far Red Wing & $12.6\pm4.9\%$  & $3.04\pm0.15\times10^{-5}$ & $3.40\pm1.49\times10^{-6}$ & 09:53:12 \\
            Far Blue Wing & $10.2\pm5.2\%$ & $2.73\pm0.14\times10^{-5}$ & $2.59\pm1.45\times10^{-6}$ & 09:54:54 \\
            \toprule
            \end{tabular}
            \label{table3}
            \end{table}
                        
            The Whole Scan band peaked earlier than the Whole Line band, likely due to the Si \textsc{iii} line exhibiting relatively stronger enhancement in the impulsive phase, while the \Ly line was relatively more enhanced during the gradual phase. In panel a.) of Figure \ref{figure4} it is shown that both the Whole Scan and Whole Line bands exhibit two distinct peaks at 09:52:56 UT and 09:59:39 UT, indicating contributions from both impulsive and gradual phase processes to \Ly enhancement. The Wing bands peaked at distinctly different times: the Near Blue Wing peaking one raster ($\sim58$s) after the Near Red Wing and Far Red Wing, while the Far Blue Wing peaked a further raster later. During the raster coinciding with the impulsive peak of the Whole Line band and HXR emission in the RHESSI $25-50\;\textrm{keV}$ channel, the Wing bands were more enhanced than the Line Core band, consistent with the findings for \flrones and previous studies \citep[e.g.][]{Brekke_1996,Woods_2004}. Panel d.) of Figure \ref{figure4} displays HXR observations in the $6-12$, $25-50$ and $50-100\;\textrm{keV}$ channels of RHESSI, alongside both GOES/XRS channels. Four HXR bursts were observed in the $25-50\;\textrm{keV}$ channel of RHESSI at 09:50:12 UT, 09:53:04 UT, 09:54:16 UT and 09:54:52 UT. The second HXR burst showed good agreement in time with peaks in \Ly enhancement in the Line Core, Near Red Wing and Far Red Wing band, which occurred during the same raster at 09:52:55 UT, 09:53:04 UT and 09:53:12 UT, respectively. The Near Blue Wing peaked at 09:54:09 UT, close to the third HXR burst, while the Far Blue Wing peaked at 09:54:33 UT, coinciding with the fourth HXR burst.
                
            Panel e.) of Figure \ref{Figure1} shows lightcurves for SORCE/SOLSTICE Whole Scan, GOES/EUVS-E and SDO/AIA $1600\;\textrm{Å}$ during \flrtwo. The timing of flare enhancement was consistent across instruments, with peak enhancements observed in GOES/EUVS-E and SDO/AIA $1600\;\textrm{Å}$ at 09:52:55 UT and 09:53:04 UT, respectively, within the same raster at which SORCE/SOLSTICE Whole Scan peaked at 09:52:55 UT. The SORCE/SOLSTICE Whole Scan recorded a peak enhancement of $5.2\%$, closely matching the peak GOES/EUVS-E enhancement of $5.8\%$, as shown in Figure \ref{figure4}. Furthermore, SDO/AIA $1600\;\textrm{Å}$ showed a comparable enhancement of $4.9\%$. As shown in panel a.) of Figure \ref{figure4}, SORCE/SOLSTICE observed significantly greater relative enhancement of the Si \textsc{iii} line of $44.4\%$ compared to the \Ly Whole Line band at $4.6\%$. The excess irradiance for Si \textsc{iii} peaked at $6.11\times10^{-5}$ Wm$^{-2}$, around one fifth the excess in the \Ly Whole Line band of $3.51\times10^{-4}$ Wm$^{-2}$. This relative contribution of the Si \textsc{iii} line to the \Ly Whole Line excess was smaller during \flrtwos than \flrone.
    
\begin{figure}[ht]    
                \centerline{\hspace*{0.015\paperwidth}
                       \includegraphics[width=.55\paperwidth,clip=]{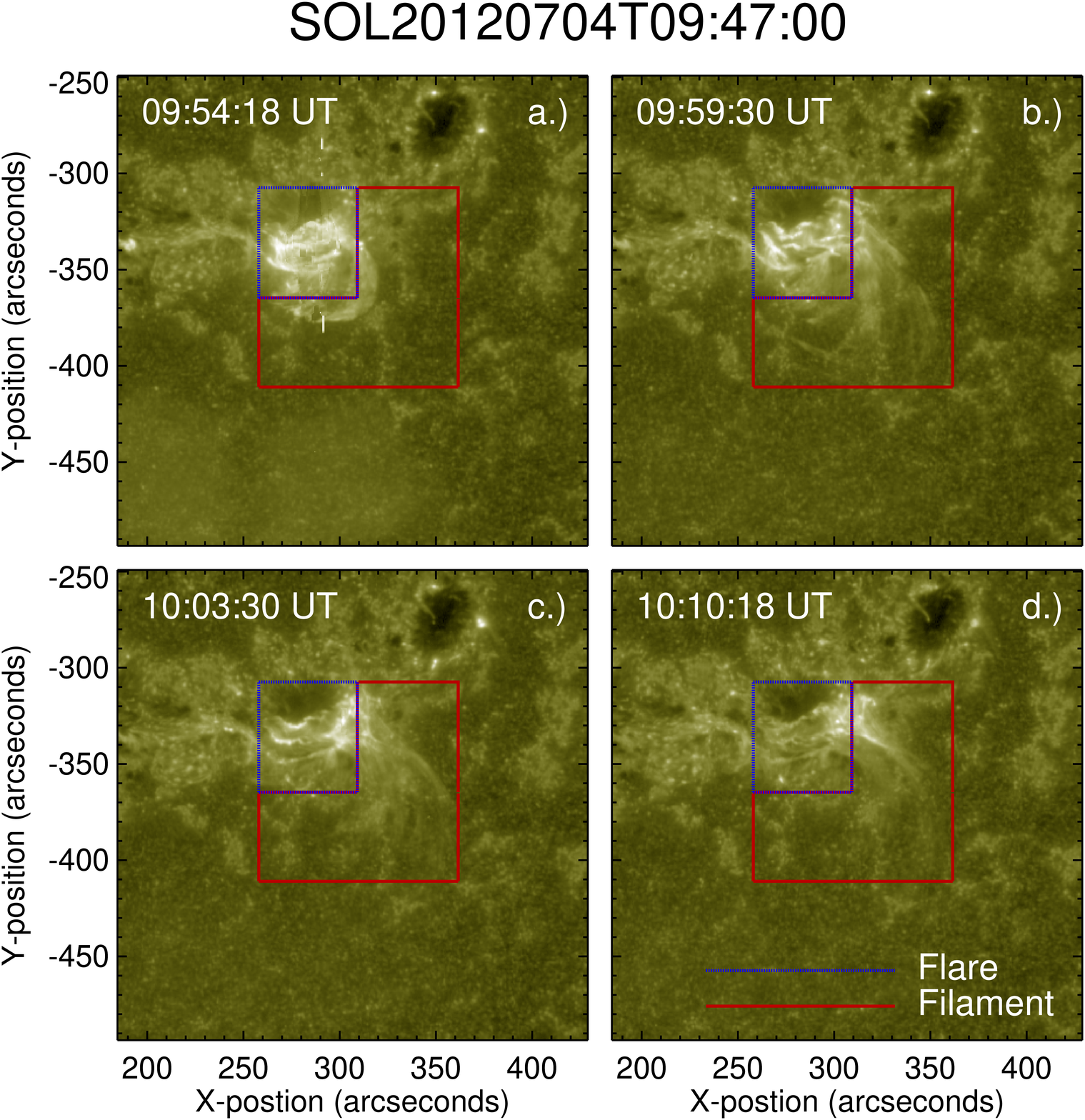}
                      }
            \caption{Desaturated SDO/AIA $1600\;\textrm{Å}$ images of \flrtwo. The flaring region is contained in the blue box while the region of filament-eruption is contained in the red box. Images shown are at times at which peaks in filament-eruption emission are seen.}
            \label{figure5}
            \end{figure}

\begin{figure}[ht]    
                \centerline{\hspace*{0.015\paperwidth}
                       \includegraphics[width=.55\paperwidth,clip=]{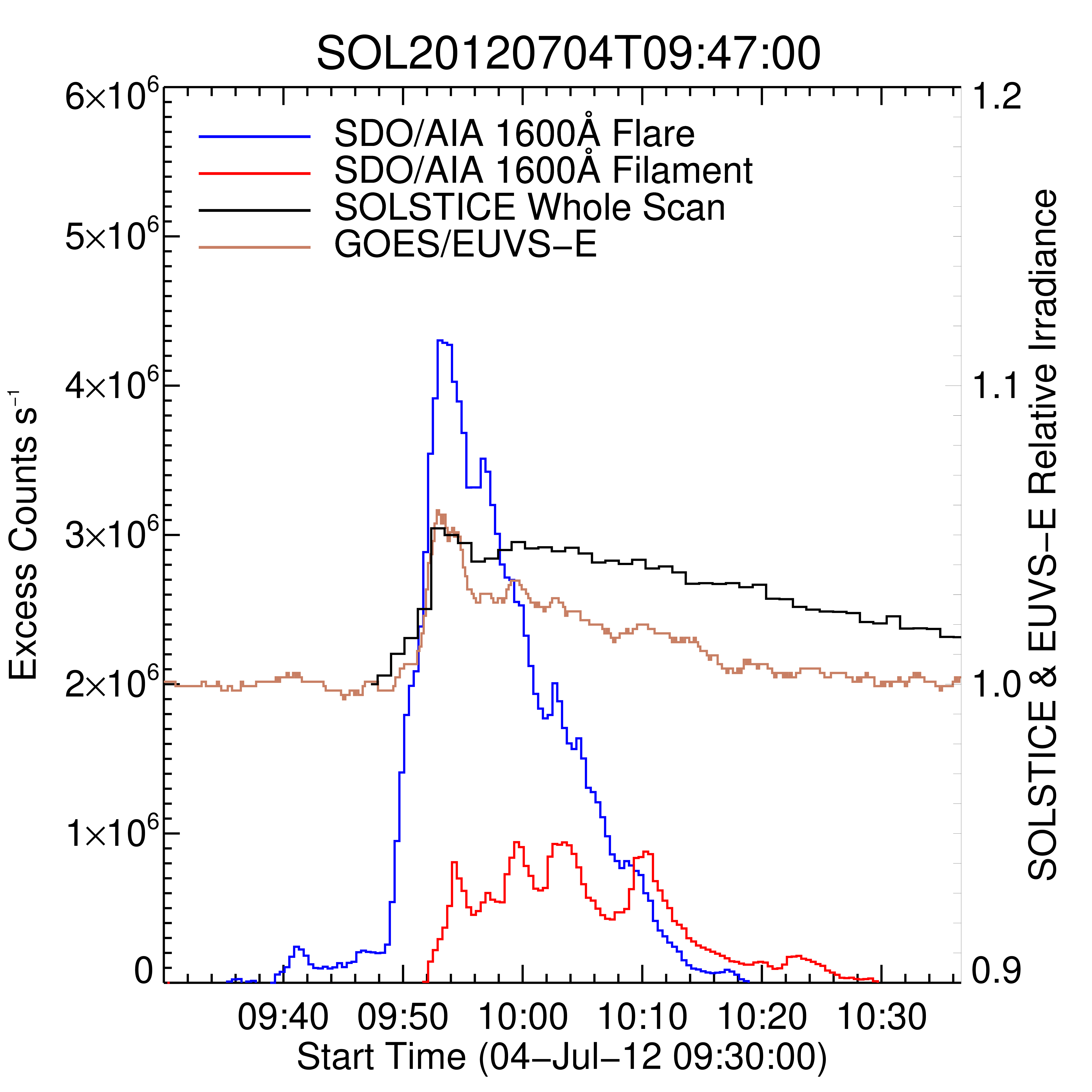}
                      }
            \caption{Lightcurves of SDO/AIA $1600\;\textrm{Å}$ excess counts for \flrtwos flare (blue) and filament (red) regions. \Ly relative irradiance is also shown for GOES/EUVS-E and SORCE/SOLSTICE Whole Scan band in brown and black, respectively. 
            }
            \label{figure6}
            \end{figure}

            Figure \ref{figure5} illustrates the division of SDO/AIA $1600\;\textrm{Å}$ images into filament and flare regions, with corresponding lightcurves shown in Figure \ref{figure6}. The flare region peaked in excess counts at 09:53:30 UT, while the filament region exhibited several distinct peaks at 09:54:18 UT, 09:59:30 UT, 10:03:30 UT and 10:10:18 UT, displayed in panels a.), b.), c.) and d.) of Figure \ref{figure5}, respectively. The first filament peak coincided with peak enhancements in the Near and Far Blue Wing bands of SORCE/SOLSTICE at 09:54:10 UT and 09:54:54 UT, respectively, suggesting that blue-shifted emission from the erupting filament contributed to these enhancements. The subsequent filament peaks correlated with the timing of gradual phase enhancements observed by GOES/EUVS-E at 09:59:06 UT, 10:02:33 UT and 10:09:39 UT, as shown in Figure \ref{figure6}, suggesting that the bright erupting filament also contributed to the gradual phase \Ly enhancements. Similar findings of filament contributions to gradual phase \Ly enhancement were reported by \citet{Wauters_2022} during an M6.7 flare using SDO/AIA $1600\;\textrm{Å}$ and PROBA-2/LYRA observations.
            
            \subsection{Comparison Between Flares}
            The comparison of \flrones and \flrtwos (Figures \ref{Figure3} and \ref{figure4}) provides valuable insights into the mechanisms driving \Ly spectral variability during flares. Both events exhibited similar peak enhancements of the Whole Scan band, with values of $5.1\%$ for \flrones and $5.2\%$ for \flrtwo. However, GOES/EUVS-E observed a significantly higher peak enhancement of $14.7\%$ during \flrone, compared to the $5.8\%$ enhancement observed during \flrtwo. Near Wing enhancements were significantly greater during \flrone, with peak values of $18.3\%$ and $15.8\%$ in the Near Red and Near Blue, respectively, compared to enhancements of $9.2\%$ and $8.5\%$ for \flrtwo. A similar trend was seen in the Far Wing bands, with \flrones exhibiting larger enhancements during of $25.4\%$ and $19.3\%$ for the Far Red and Far Blue bands, respectively, compared to $12.6\%$ and $10.2\%$ during \flrtwo. The Si \textsc{iii} line also showed significantly greater peak enhancement of $97.5\%$ during \flrone, in contrast to the $44.4\%$ for \flrtwo. For both events, enhancements in each SORCE/SOLSTICE band were temporally correlated with bursts of HXR emission, indicating these enhancements were likely driven by nonthermal processes. Additionally, enhancements in the Near and Far Blue bands of SOLSTICE, as well as in GOES/EUVS-E during \flrtwo, were attributed to a bright filament-eruption observed by SDO/AIA. \comment{} Blue wing enhancements associated with filament eruptions during flares have also been identified in sun-as-a-star observations from SDO/EVE, in various other chromospheric and transition region lines such as Ly$\beta$, He \textsc{i} $584.3\;\textrm{Å}$ and O \textsc{v} $629.7\;\textrm{Å}$ \citep{Brown_2016, Xu_2022,Lu_2023,Otsu_2024}\commentend{}. Although STEREO/SECCHI EUVI images revealed a filament-eruption during \flrone, it was not possible to compare its emission to that in \Ly due to significant saturation and the low cadence of the observations. As both events had similar GOES class, it is likely that varied contributions of flare processes such as nonthermal heating, conduction and radiative cooling between the flares drove the observed differences in \Ly variability. 

\section{Discussion \& Conclusions}
    \label{S- Conclusions}
    This study analysed the spectral irradiance variability of \Ly emission during two M-class flares, \flrones and \flrtwo, using newly-released high-cadence calibration scans from SORCE/SOLSTICE. By comparing \Ly spectral enhancements with HXR and UV observations, the heating mechanisms responsible for the observed variability were investigated. During \flrone, good agreement was observed between HXR emission and enhancements across the \Ly line profile, suggesting that nonthermal heating primarily drove these \Ly enhancements. Additionally, during \flrtwo, enhancements in the Line Core and Red Wing bands cotemporal with a HXR burst further supported a nonthermal origin for the spectral variability. These findings are consistent with previous studies that compared HXR emission and enhancement of other chromospheric lines such as H$\alpha$ \citep{Kurokawa_1988, Radziszewksi_2011}, supporting the role of nonthermal electrons in driving chromospheric \Ly enhancement during flares. The Blue Wing bands during \flrtwos exhibited peak enhancements at a later time, coincident with a further HXR burst and emission from a bright filament-eruption seen in SDO/AIA $1600\;\textrm{Å}$ images. This overlap makes it unclear whether the Blue Wing enhancements were primarily driven by nonthermal chromospheric heating or by filament-related emission. Furthermore, gradual phase filament emission correlated with \Ly enhancements observed by GOES/EUVS-E, supporting findings from previous studies \citep{Milligan_2021,Wauters_2022}. These results provide further evidence that filament-eruptions likely contribute to \Ly flare variability along with traditional flare processes such as nonthermal heating, conduction and radiative cooling.
    
    Both flares exhibited red asymmetry in wing enhancements, likely indicative of chromospheric evaporation \citep{Hong_2019}. During \flrtwo, the red enhancement asymmetry changed to blue, coinciding with emission from a filament eruption. This suggests the filament may have contributed to the observed blue asymmetry via blue-shifted emission. While previous observational and theoretical studies have mostly focused on \Ly emission driven by heating of the chromosphere by nonthermal particles and thermal conduction, future research should also explore the potential impact of emission from filament eruptions on spectral variability during flares.
    
    Comparisons between GOES/EUVS-E and SORCE/SOLSTICE revealed marked differences in \Ly enhancement during \flrone, while better agreement was found between the instruments during \flrtwo. These discrepancies highlight ongoing challenges in \Ly flare observations, with different instruments providing disparate results for the same events \citep[e.g.][]{Milligan_2016,Greatorex_2024}. Future comparisons of SORCE/SOLSTICE observations with other \Ly instruments may help uncover the reasons for these disparities.
      
    The key conclusion from this study is that \Ly spectral variability during flares is likely primarily driven by nonthermal processes in the chromosphere but may also be influenced by coronal structures, such as filament-eruptions. Comparisons between SORCE/SOLSTICE observations and simulated \Ly flare line profiles may further constrain the results of the simulations, providing further insight on the mechanisms driving \Ly spectral variability \citep{Brown_2018,Kerr_2023}. Additionally, SORCE/SOLSTICE observations may provide insight into \Ly variability during stellar flares and the influence of flaring \Ly on exoplanet atmospheres \citep{Lecavelier_des_Etangs_2012,Hazra_2022}. 

    Despite its contributions, this study faced several limitations. The small number of SORCE/SOLSTICE flare observations restricted the statistical scope of this analysis, with only two flares being studied. Additionally, the rastering nature of SOLSTICE, combined with its modest wavelength resolution ($0.35\;\textrm{Å}$) and substantial irradiance uncertainty hampered the interpretation of \Ly spectral variability. As a result, the analysed flares, which had uncharacteristically large \Ly enhancements, may not reflect the spectral variability of weaker flares.
    
    Upcoming instruments, such as the EUV High-Throughput Spectroscopic Telescope aboard the SOLAR-C mission \citep[EUVST;][]{Watanabe_2014,Shimizu_2019}, are expected to provide state-of-the-art, spectrally resolved \Ly flare observations. EUVST will feature a dramatically higher wavelength resolution ($0.008\;\textrm{Å}$) and cadence ($1\;\textrm{s}$) than SOLSTICE. This will eliminate any asymmetries caused by rastering and enable a more precise interpretation of the flare dynamics driving spectral variability by resolving spatial features in the chromosphere and corona. Additionally, the Solar eruptioN Integral Field Spectrograph (SNIFS) will provide high wavelength resolution ($0.0033\;\textrm{Å}$) and cadence ($1\;\textrm{s}$) observations of \Ly, potentially capturing a flare during its sounding rocket flight. These instruments should combine to provide a wealth of high-quality flare observations, enabling the detection of smaller \Ly enhancements than previously possible. This will facilitate both statistical and qualitative studies, offering further insights into flare dynamics in the chromosphere and corona. 
    
    Combined \Ly spectral and HXR observations from EUVST, SNIFS, Solar Orbiter's Spectrometer Telescope for Imaging X-rays \citep[SolO/STIX;][]{Muller_2020,Krucker_2020}, and the Hard X-ray Imager aboard the Advanced Space-Based Solar Observatory \citep[ASO-S/HXI;][]{Zhang_2019} will enable a more conclusive determination of the relationship between \Ly spectral variability and the properties of nonthermal electrons that drive this variability. These observations will benefit from true \Ly imaging from EUVST, SNIFS, SolO/EUI and Lyman Solar Telescope onboard ASO-S \citep[ASO-S/LST][]{Gan_2019,Li_2019}. The combination of spectral and imaging data will enable the disentanglement of chromospheric and coronal contributions to \Ly spectral variability. Furthermore, HXR observations will enable the comparison of spatial sources of \Ly and HXR emission, revealing whether \Ly emission in different features is driven by nonthermal heating. This analysis will provide a deeper understanding of the physical processes driving \Ly spectral variability during flares. 
    
\begin{acks}
L.H.M. acknowledges support from the Department for the Economy (DfE) Northern Ireland postgraduate studentship scheme. R.O.M. and E.C.B. acknowledge support from STFC grants ST/W001144/1 and ST/X000923/1.
\end{acks}

\bibliographystyle{spr-mp-sola}
\bibliography{bibliography.bib} 

\end{article} 

\end{document}